\documentclass[prb,aps,twocolumn,amsmath,amssymb,amsfonts,floatfix,showkeys,showpacs,superscriptaddress]{revtex4-1}%,
\usepackage{dcolumn}  % Align table columns on decimal point
\usepackage{multirow}
\usepackage[T1]{fontenc}
\usepackage{dsfont}
\usepackage{verbatim}
\usepackage{bm} % bold math 
\usepackage{hyperref} 
\usepackage{cleveref} 
\usepackage{floatrow} 
\usepackage{natbib}
\usepackage{graphicx}
\usepackage{epstopdf}
\usepackage{ifpdf} % This lets us get figures right both in pdf and ps versions
\usepackage{epsfig}
\usepackage{color}
\usepackage[usenames,dvipsnames]{xcolor}
\usepackage{enumitem}

\newcommand{\icomplex}{\dot\iota}%other notation: \dot\iota, \imath, \dot\imath, {\mathrm{i}}
\usepackage{scalerel}

\begin{document}

\title{‌Bag boundaries for quasispinor confinement within nanolanes on a  graphene sheet}

\author{Yusef Koohsarian}\email{koohsarian.ramian@gmail.com (corresponding author)}
\affiliation{School of Nano Science, Institute for Research in Fundamental Sciences (IPM), Tehran 19538-33511, Iran}
\affiliation{School of Physics, Institute for Research in Fundamental Sciences (IPM), Tehran 19538-33511, Iran}
\author{Ali Naji}\email{a.naji@ipm.ir}
\affiliation{School of Nano Science, Institute for Research in Fundamental Sciences (IPM), Tehran 19538-33511, Iran}
\affiliation{Department of Physics, College of Science, Sultan Qaboos University, Muscat 123, Oman}
 
%%%%%%%%%%%%%%%%%%%%%%%%%%%%%%%%%%%%%%%%%%%%%%
\begin{abstract}
 We revisit the problem of bag boundary conditions within a field-theoretic approach to study   confinement of massless Dirac quasispinors  in monolayer graphene. While no-flux bag boundaries have previously been used to model lattice termination sites in graphene nanoribbons, we consider a generalized setting in which the confining boundaries are envisaged as arbitrary straight lines drawn across a graphene sheet and the quasispinor currents are allowed to partially permeate (leak) through such boundaries. We specifically focus on rectangular nanolanes defined as areas confined between a pair of parallel lines at arbitrary separation on an unbounded lattice. We show that such nanolanes exhibit a considerable range of bandgap tunability depending on their widths and  armchair, zigzag or intermediate orientation.  The case of nanoribbons can be derived as a special limit from the nanolane model. In this case, we clarify certain inconsistencies in previous implementations of no-flux bag boundaries and show that the continuum approach reproduces the tight-binding bandgaps accurately (within just a few percent in relative deviation) even as the nanoribbon width is decreased to just a couple of lattice spacings. This accentuates the proper use of boundary conditions when field-theoretic approaches are applied to graphene systems.
\end{abstract}

\maketitle
 
%%%%%%%%%%%%%%%%%%%%%%%%%%%%%%%%%%%%%%%%%%%%%%
\section{Introduction}   

The discovery of graphene---the prominent single-layer honeycomb lattice of carbon atoms \cite{G1,G2,G3,G4}---has made a significant impact on a wide range of modern  technological applications  \cite{Gtec1,Gtec2,Gtec3,Gtec4,Gtec5}. The advances are made possible thanks to many remarkable transport, optical and mechanical properties that have since been uncovered for graphene and its derivatives. One of the key physical aspects of monolayer graphene is its gapless spectra of low-energy (electron-hole) excitations that emerge as massless Dirac quasispinors \cite{GD,GR1,GR2,GR3,GR4,GR5}. This brings the low-energy physics of graphene into close analogy with the quantum field theory of Dirac spinors. Massless Dirac quasispinors are responsible for some of the most exotic phenomena observed in  graphene such as anomalous quantum Hall effect \cite{G-H1,G-H2,G-H3} and Klein tunneling \cite{G-K1,G-K2,G-K3}. 
%The latter itself remains a long-standing prediction for Dirac spinors in quantum field theory \cite{Klein1,Klein2}. 
These have driven a substantial interest in field-theoretic formulations of graphene physics in the recent past  \cite{GR1,GR2,GR3,GR4,GR5,GQFTrev1,GQFTrev2,GQFT1,GQFT2,GQFT3,GQFT4}.  

While unbounded graphene has thoroughly been studied using field-theoretic methods, its other realizations such as graphene nanoribbons and carbon nanotubes have received less attention from that perspective \cite{NR5,NR6,NR7,GQFT1,GQFT2,GQFT3}. Graphene nanoribbons have emerged as  promising semiconducting materials with tunable energy bandgap in room-temperature nanoelectronic applications  \cite{Gappl1,Gappl2,Gappl3}. The most commonly studied cases of nanoribbons involve a regular, armchair or zigzag, arrangement of carbon atoms on their side edges. The bandgap of fixed-width nanoribbons can be calculated  using standard tight-binding methods and is found to depend on the nanoribbon width and edge configuration \cite{NR1,NR2,NR3,NR4}.  The presence of side edges raises the subtle issue of boundary conditions when a field-theoretic formulation is attempted. In the tight-binding approach, the boundaries are modeled by imposing Dirichlet-type boundary conditions on electronic wavefunctions at nanoribbon side edges  \cite{NR4}. In the field-theoretic approach, Dirichlet-type boundary conditions are useless due to their well-known inconsistency with Dirac equations. Instead, bag boundary conditions have recently been utilized to model vanishing quasispinor currents at the lattice termination sites  \cite{GQFT2,GQFT3,GQFT4}.  

Bag models were originally proposed in quantum chromodynamics as a consistent route to confine Dirac spinors (quarks in hadrons) by imposing  no-flux conditions along surface normals of an assumed confining volume (resembling an impermeable `bag')  \cite{bag,bag1,bag2,bag3,bag3b,bag4,bag5,bag6}.  For armchair graphene nanoribbons, the consistency issue was also addressed    \cite{NR5,NR6} via a particular type of boundary condition that  admixed the so-called valleys of the energy band at armchair side edges. It was later shown  \cite{GQFT3}  that the proposed boundary condition is  equivalent to the no-flux bag model   even as sizable differences in predicted bandgaps  persisted  \cite{NR5} relative to tight-binding results. 

In this paper, we revisit the problem of bag boundaries for quasispinor confinement in graphene by placing it in a broader context beyond its prior application to nanoribbons. Rather than considering them as lattice termination sites as done for nanoribbon,  we envisage confining boundaries as  arbitrary in-plane lines  introduced  across an unbounded graphene sheet. The specific case of interest in this work will be a pair of  parallel straight lines that confine a quasi-one-dimensional (quasi-1D) area over the graphene sheet which we refer to as a {\em nanolane}. Hence, in further contrast to nanoribbons, the width of nanolanes is (by construction) treated as a {\em continuous} variable. The boundary orientation can also be varied  (in analogy with nanoribbons) to define  armchair, zigzag and intermediate nanolanes. Taking advantage of the continuum field-theoretic description, we introduce a generalized form of bag boundaries  by allowing finite fluxes of quasispinors to  permeate or `leak' through the boundary of a nanolane.

Since the formulation of the current model is rather generic, its predictions could  be relevant  potentially to a host of real situations involving partial (`leaky') electronic confinement in graphene. For instance, the  boundary lines can be viewed as arrangements of adatoms or, when a graphene sheet is suspended over a rectangular trench, as the parallel lines of contact between the sheet and the edges of the trench. In the latter case, the lines of contact may not necessarily match   armchair/zigzag edges of a respective nanoribbon and may also allow for quasispinor permeation,  making the proposed nanolane model a more suitable modeling alternative to consider.  

The quantitative viability of our model is supported by comparing its predictions in the special limit of nanoribbons (i.e., nanolanes with no-flux boundary lines and discrete widths) with the tight-binding results    \cite{NR1,NR2,NR3,NR4}. In contrast to Ref.  \cite{NR5}, where nanoribbon bandgaps are studied through no-flux bag boundaries and sizable deviations from the tight-binding bandgaps at small widths are reported,  our model reproduces the  tight-binding results within  relative differences of just a few percent even at small widths around a couple of lattice spacings. This signifies the extended applicability of the continuum field-theoretic approach in the present context. 

 In Section \ref{subsec:intro}, we provide a brief background on Dirac quasispinors and discuss the details of our model whose predictions for armchair, zigzag and intermediate nanolanes are then given in Sections \ref{subsec:acGNL}-\ref{subsec:int}. A comparison with the case of nanoribbons is given in Section \ref{subsec:GNR}, followed by the concluding remarks in Section \ref{sec:conc}.

\begin{figure}[t!]
\centering
\includegraphics[width=\textwidth]{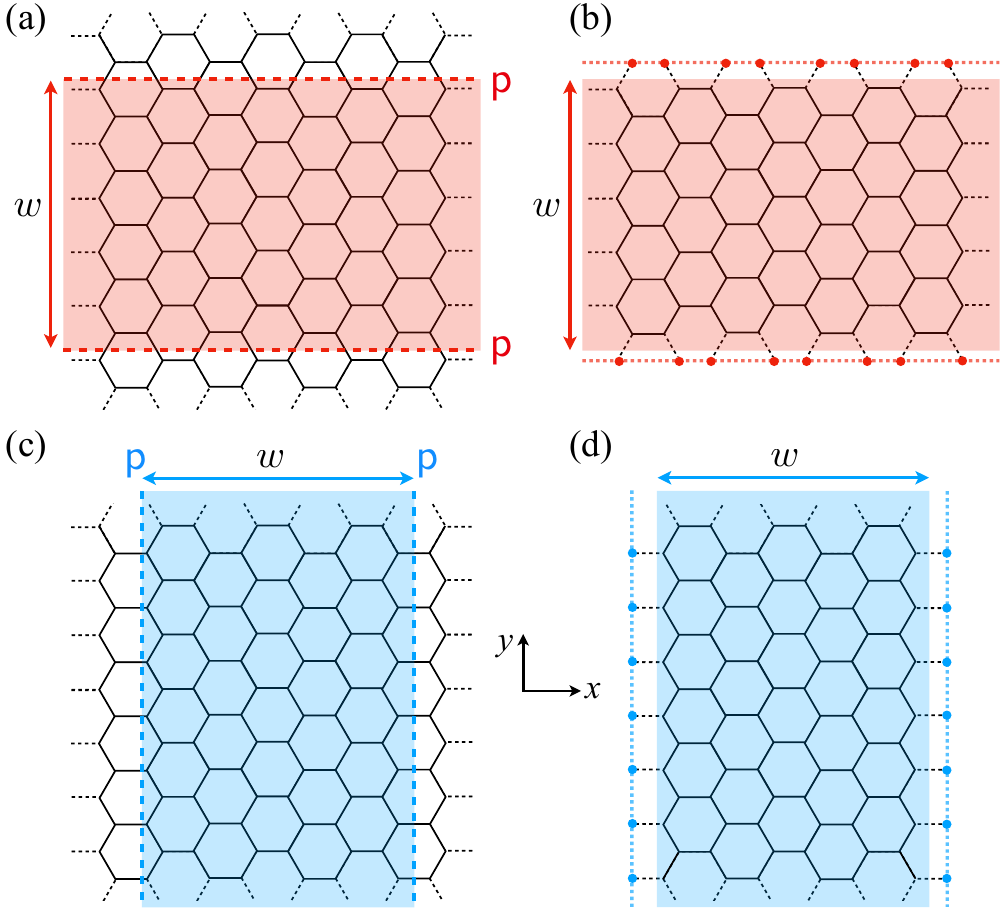}
\caption{Armchair (a) and zigzag (c) nanolanes  (colored areas) are defined by envisaging in-plane boundary lines (dashed lines) with finite permeability $\mathsf{p}$ relative to quasispinor currents over an otherwise infinite honeycomb lattice. Armchair (b) and zigzag (d) nanoribbons can be recovered as special limits of nanolanes by setting the permeabilities equal  to zero ($\mathsf{p}=0$) and  shifting the boundaries (dotted lines) by a total offset of $d$ to define lattice termination sites.  In (a) and (c), the dotted bond lines on the margins of the two schematics indicate infinite continuation of the honeycomb lattice. In (b) and (d), the red/blue bullets on the margins indicate  passivating adatoms  \cite{NR2} (and not carbon atoms themselves as one would encounter in bearded nanoribbons).}
\label{GNLs}
\end{figure}

%%%%%%%%%%%%%%%%%%%%%%%%%%%%%%%%%%%%%%%%%%%%%%
\section{Model description}   
\label{subsec:intro}

%%%%%%%%%%%%%%%%%%%%%%%%%%
\subsection{Preliminaries on Dirac quasispinors in graphene}
\label{subsec:preliminaries}

\begin{figure}[b!]
\centering
\includegraphics[width=3.25cm]{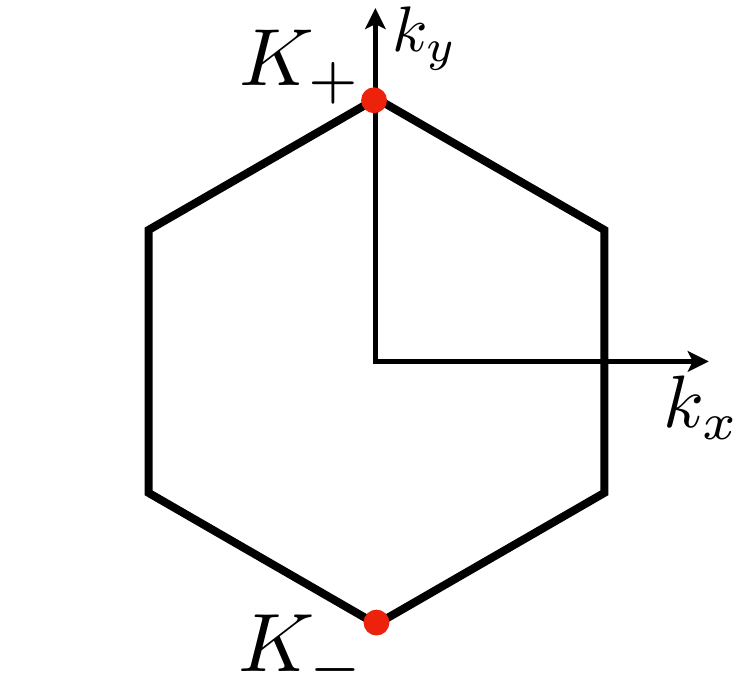}
\caption{ First Brillouin zone of graphene with the independent Dirac points taken as $\mathbf{K}_\pm=(0,\pm\frac{4\pi}{3d})$ (solid red circles).}
\label{BZ}
\end{figure}

Quasispinor representation of electronic states in an unbounded monolayer graphene follows standardly from a tight-binding formulation of in-plane $\pi$-electron hopping between nearest-neighbor carbon atoms. Thanks to its two-dimensional (2D) honeycomb lattice (incorporating two triangular sublattices), an undoped monolayer displays conduction and valence energy bands in the 2D reciprocal  space that meet at six individual Dirac points at the corners of the first Brillouin zone.  Only two of the Dirac points turn out to be independent \cite{GR1}. They can suitably be chosen as $\mathbf{K}_\pm=(0,\pm\frac{4\pi}{3d})$,  see Fig. \ref{BZ},  where $d =  \sqrt{3}  d_0$  is the lattice spacing and $d_0\simeq 1.42$\,\AA\, is the carbon-carbon bond length.

 Expanding the tight-binding Hamiltonian near the Dirac points $\mathbf{K}_\pm$ up to the first order in the relative wavevector $\mathbf{q}= (q_1,q_2)= \mathbf{k}-\mathbf{K}_\pm$ leads to pseudo-Dirac Hamiltonians \cite{GR1}
\begin{eqnarray}
 H_{\pm, \mathbf{q}}&=& \hbar \upsilon_{\mathrm{F}}  \begin{pmatrix} 0 & \icomplex q_1\pm q_2 \\ - \icomplex q_1\pm q_2  &0 \end{pmatrix}, 
%\nonumber\\
% &=&\hbar\upsilon_{\mathrm{F}}(-\sigma_2 q_1\pm \sigma_1q_2),
 \label{2H}
\end{eqnarray}
where $\upsilon_{\mathrm{F}} \equiv 3 t d_0/2\simeq 10^6\,\mathrm{m}/\mathrm{s}$  is the  Fermi velocity and $t$ is the nearest-neighbor hoping parameter of the underlying tight-binding model. The Hamiltonian \eqref{2H} acts on the 2-spinor fields $\psi_{\pm,\mathbf{q}}$ whose components correspond to probability amplitudes of Bloch wavefunctions on the two underlying sublattices of graphene at each Dirac point. By introducing the 4-spinor field $  \Psi_{\mathbf{q}}  =\left(\psi_{+,\mathbf{q}},\psi_{-,\mathbf{q}}\right)^T$, the two blocks $H_\pm$ can be combined into a single Hamiltonian as  $H=\icomplex\hbar \upsilon_{\mathrm{F}} \gamma ^0\gamma^a\partial_a$ where Einstein's convention is used for summation over $a=1,2$ and $q_a$ is replaced by $ -\icomplex \partial_a \!\equiv\! -\icomplex \partial/\partial x_a$. Hereafter, the spatial coordinates are  denoted interchangeably  by $(x_1,x_2)$ or $(x,y)$. We have also used the definitions \cite{GQFT1,GQFT2,GQFT3,GQFT4,GQFTrev1,GQFTrev2} 
\begin{equation}
 \gamma^0= \sigma_3 \otimes {\mathbb I},\ \  \gamma^{1}= \icomplex \sigma_1 \otimes {\mathbb I} , \ \  \gamma^{2}= \icomplex \sigma_2 \otimes \sigma_3, 
 \label{gam}
\end{equation}
where $\sigma_1$, $\sigma_2$ and $\sigma_3$ are standard Pauli matrices and ${\mathbb I}$ the is the unit matrix. The aforesaid Hamiltonian leads to the massless pseudo-Dirac equation 
 \begin{equation}
\icomplex  \upsilon_{\mathrm{F}}\gamma^0\gamma^a\partial_a\Psi_\mathbf{q}(\mathbf{r}) =  \omega_\mathbf{q} \Psi_\mathbf{q}(\mathbf{r}),   
\label{4H}
\end{equation}
where $\omega_\mathbf{q}= \upsilon_{\mathrm{F}} |\mathbf{q}|$. It is useful to note that the dispersion relation itself adopts positive and negative  solutions $\epsilon_\mathbf{q}= \lambda \hbar\omega_\mathbf{q}$   identified by the conduction versus valence band index $\lambda=\pm 1$ (this reflects the sublattice or quasispin index, which is not explicitly indicated here, and  should not be confused with the $\pm$ subscripts used  to denote the valley index associated with the two Dirac points $\mathbf{K}_\pm$). 

For brevity, we henceforth drop the subscript $\mathbf{q}$. We also note in passing that the spatial quasispinor currents $J^{a}=\Psi^\dag \gamma^0 \gamma^a \Psi$ can be written as  $J^{a}=J^{a}_++J^{a}_-$ where each of the terms 
 \begin{equation}
   J^{a}_\pm=\psi_\pm^\dag \icomplex\sigma_3\sigma_a\psi_\pm\quad (a=1,2)
   \label{J^a}
   \end{equation} 
incorporate both the electron and hole contributions. 

%%%%%%%%%%%%%%%%%%%%%%%%%%
\subsection{Nanolanes with permeable bag boundaries}
\label{subsec:bag_BC}

In principle, any external source that could break the isotropy of the $\pi$-electron distribution in monolayer graphene would produce an in-plane boundary condition on quasispinors.  The permeable bag boundary conditions that will be of primary interest here are imposed over (in-plane) quasispinor currents across contour lines prescribed over an infinite graphene sheet. We shall consider two such straight lines drawn in parallel across the honeycomb lattice. These  boundary lines confine a narrow quasi-1D region of the underlying lattice in between that we refer to as a {\em nanolane}. The cases of armchair and zigzag nanolanes follow when the boundary lines are set parallel to the armchair bonds and zigzag vertices, respectively. These are visualized in Figs. \ref{GNLs}a and c. In the schematics,  the rectangular areas colored in light red and blue show the nanolane regions and the thick dashed same-color  lines indicate the imposed  boundary lines (see also Fig. \ref{im-e} for the intermediate case of a tilted nanolane).

 The permeable bag boundary conditions at a given boundary line reads 
 \begin{equation}
\mathbf{J}_\pm\cdot {\mathbf{n}} = \mathsf{p}_\pm, 
 \label{BB}
 \end{equation} 
where   $\mathsf{p}_\pm$ are the {\em permeabilities} of the boundary in question relative to quasispinor currents associated with $\psi_\pm$. Here, ${\mathbf{n}}=(n^1, n^2)$ denotes the inward  unit  boundary normal and $\mathbf{J}_\pm=(J^{1}_\pm,J^{2}_\pm)$ the spatial quasispinor currents with spatial components given in Eq. \eqref{J^a}.

 To obtain an {\em armchair nanolane} of width $w$, we set the boundary positions at $y=\mp w/2$, giving the corresponding boundary normals as ${\mathbf{n}}=(0, \pm 1)$. For a {\em zigzag  nanolane}  of width $w$, we set the boundaries at $x=\mp w/2$ and,  hence, the boundary normals are ${\mathbf{n}}=(\pm 1, 0)$.

 For further succinctness, we drop the subscripts $\pm$ from $\psi_\pm$ and $\mathsf{p}_\pm$ since (because of the  block-diagonal form of the Hamiltonian) the upcoming calculations can similarly be repeated for either of the two quasispinor fields. We also focus on the case of {\em symmetric nanolanes} for which  the permeability coefficients at the two boundaries are equal (Figs. \ref{GNLs}a and c). In general, the boundary permeability can vary with the wavevector $ \mathbf{q}$. However, as we shall see later, the exact form of the permeability coefficient does not affect the energy spectrum for symmetric nanolanes.
 
 Before proceeding further, we emphasize that (1) the nanolane construction  involves no bond `cuts' in the background (infinite) honeycomb lattice, and (2) by definition, the nanolane width $w$ can be varied continuously. These properties distinguish nanolanes from {\em nanoribbons}. In the latter case,  the boundary lines are actual edges or lattice termination points (e.g., with saturated bonds) and the width can only vary over a discrete set of values (see Figs. \ref{GNLs}b and d).  
%From a practical perspective, the nanolane model can thus be more relevant to other possible realizations of graphene; e.g., a suspended sheet over a rectangular trench (see the discussion in Section \ref{sec:conc}).
Yet, despite these differences, the present model can formally reproduce the case of nanoribbons as a special limit of nanolanes; i.e., upon setting the boundary permeabilities to zero $\mathsf{p}=0$ and shifting the boundary lines by a total offset of $d$ to position them at their relevant termination sites (Figs. \ref{GNLs}b and d).  The no-flux (or impermeable) form of bag boundary conditions has indeed been applied to model graphene nanoribbons before \cite{GQFT2,GQFT3,GQFT4,NR5,NR6} but with sizable departures from the known tight-binding results \cite{NR1,NR2,NR3,NR4} which we address and clarify  in Section \ref{subsec:GNR}.

%%%%%%%%%%%%%%%%%%%%%%%%%%%%%%%%%%%%%%%%%%%%%%
\section{Results}   

%%%%%%%%%%%%%%%%%%%%%%%%%%
\subsection{Armchair nanolanes}
\label{subsec:acGNL}
 For an armchair nanolane,  Eq. \eqref{4H} can be solved inside the nanolane region ($|y|<w/2$). The generic form of solutions for the 2-spinor field are expectedly similar to those of nanoribbons (see, e.g., Refs. \cite{GR1,NR5,NR6,GQFT2,GQFT3}); i.e.,  
  \begin{equation}
 \psi_\mathbf{q}(x,y) = \frac{{\mathrm{e}}^{\icomplex q_1 x}}{\sqrt{2}}\Big[A\begin{pmatrix} {\mathrm{e}}^{-\icomplex \theta/2 }\\ {\mathrm{e}}^{\icomplex \theta/2 }\end{pmatrix} {\mathrm{e}}^{\icomplex q_2 y}+B\begin{pmatrix} {\mathrm{e}}^{\icomplex \theta/2 }\\ {-\mathrm{e}}^{-\icomplex \theta/2 }\end{pmatrix} {\mathrm{e}}^{-\icomplex q_2 y}\Big], 
  \label{psi}
  \end{equation}
where  $A$ and $B$ are  constants  and $\theta =\tan^{-1}(q_1/q_2) $. The boundary conditions can be imposed through Eq. \eqref{BB} at the  top/bottom boundaries (Fig. \ref{GNLs}a). Hence, with the appropriate choice of boundary normals (Section \ref{subsec:bag_BC}) and the relation $\sigma_3 \sigma_2=- \icomplex \sigma_1$,  we have 
  \begin{equation}
  \psi^\dag \sigma_2\psi \big|_{y=\mp w/2}=\textsf{p}
  \label{acBB}
  \end{equation}
at the bottom and  top  boundaries, respectively.  Using Eqs. \eqref{psi} and \eqref{acBB}, we find $ \sin  (q_2 w)  =  0$ which gives the permissible values of the wavevector component in $y$-direction, $q_{2,n}=n \pi/w$ where $ n$ is an integer. For the original wavevectors $\mathbf{k}=(k_1,k_2)$ in the first Brillouin zone,  $k_1 = q_1$ thus varies over the real axis, while $k_{2,n} = {4\pi}/(3d)+q_{2,n}$ takes the discrete set of values 
   \begin{equation} 
    k_{2,n}=\frac{\pi}{w}\left[n+\frac{4}{3}\!\left(\frac{w}{d}\right)\right]. 
  \label{k-ac}
  \end{equation}
 The low-energy spectra of an armchair nanolane then follows as $ \epsilon_n(k_1)= \pm \hbar \upsilon_\textrm{F}\left(k_1^2+ k_{2,n}^2\right)^{1/2}$   (with $\pm$ signs here accounting for the conduction and valence energy bands; see Section \ref{subsec:preliminaries}). Hence,  the energy bandgap is found as 
   \begin{equation}
\!\frac{\epsilon_\text{g}}{\hbar \upsilon_\textrm{F}}\!=\!\frac{2\pi}{w}\cdot\left\{
 \begin{array}{lll}
  \left|\Lambda-\ell\right|   \,\,\,\, & : &  \,  0 \leq \Lambda -\ell <\frac{1}{2},  \\ \\
  \left|\Lambda-\ell-1\right| \,\,\,\, & : & \,   \frac{1}{2} \leq \Lambda-\ell< 1, 
\end{array}
\right.
\label{ep-ac}
\end{equation}
for nonnegative integer $\ell=0, 1, 2,\ldots$ and $ \Lambda\equiv  4w/3d$.  Needless to say that expressions \eqref{ep-ac} represents the minimum of energy gaps for a given $\ell$. For instance, taking $\ell=0$, the  minimum gap for  $ 0 \leq \Lambda <{1}/{2}$ is given by
$\epsilon_\text{g}= (2\pi \hbar \upsilon_\textrm{F})\Lambda/w$, while for $ {1}/{2} \leq \Lambda< 1$, it is $\epsilon_\text{g}= (2\pi \hbar \upsilon_\textrm{F}) |\Lambda-1|/w$, etc.

\begin{figure}[t!]
\centering
\includegraphics[width=0.9\textwidth]{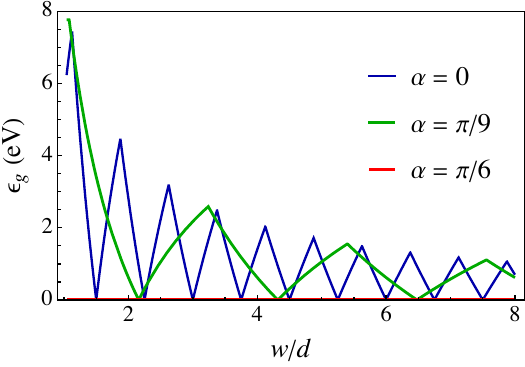}
\caption{ Solid blue line shows the energy bandgap, $\epsilon_\text{g}$, of armchair nanolanes as a function of rescaled nanolane width, $w/d$, as obtained from Eq.  \eqref{ep-ac}. Solid red line shows the vanishing bandgap of zigzag nanolanes (Section \ref{subsec:zzGNL}).  Solid green curve shows that of an intermediate nanolane with tilt angle $\alpha=\pi/9$ between those of an armchair ($\alpha=0$) and zigzag nanolane ($\alpha=\pi/6$); see Section \ref{subsec:int}.}
\label{ac-e}
\end{figure}

The gap function $\epsilon_\text{g}$ for an armchair nanolane is shown in Fig. \ref{ac-e}, solid blue line,  as a function of the nanolane width $w$. As dictated by  Eq. \eqref{ep-ac}, $\epsilon_\text{g}$ is a continuous piecewise linear function of $w$ with specific peaks and troughs. The peak values of $\epsilon_\text{g}$ vary over a wide range (dropping from $\epsilon_\text{g}\simeq 8$~eV to sub-eV values) as the nanolane width is increased modestly (from $w/d\sim 1$). The bandgap vanishes at the troughs that are obtained at   $w_\ast = (3d/4)\ell $ for $\ell=1, 2, 3, \ldots$, as visualized in Fig. \ref{ac-e}. These particular widths identify the {\em metallic states} of armchair nanolanes. 

%%%%%%%%%%%%%%%%%%%%%%%%%%%%
\subsection{Zigzag naolanes}
\label{subsec:zzGNL}

 For a zigzag nanolane, Eq. \eqref{BB} can be written at the left  and right   boundaries (Fig. \ref{GNLs}c)  as  
   \begin{equation}
  \psi^\dag \sigma_1\psi \big|_{x=\mp w/2}=\textsf{p}.
  \label{zzBB}
  \end{equation}
This can be used with the general form of the solution  \eqref{psi} to obtain the condition $ \sin  (q_1 w)  =  0$ on the permissible values of the wavevector component in $x$-direction, $q_{1,n}=n \pi/w$ where $ n$ is an integer.  Hence,  $k_{1,n} =  q_{1,n}$ takes the discrete set of values 
\begin{equation}
 k_{1,n}=\frac{n \pi}{w}, 
 \label{k-zz}
 \end{equation}
while $k_2 = q_2$   can vary over the real axis.  The low-energy spectra  $ \epsilon_n(k_1)= \pm \hbar \upsilon_\textrm{F}\left(k_1^2+ k_{2,n}^2\right)^{1/2}$ thus show vanishing bandgap $\epsilon_\text{g}=0$. In other words, the zigzag nanolanes are always metallic regardless of their width.  This case corresponds to the solid red line in Fig. \ref{ac-e}.

\begin{figure}[t!]
\centering
\includegraphics[width= 0.85\textwidth]{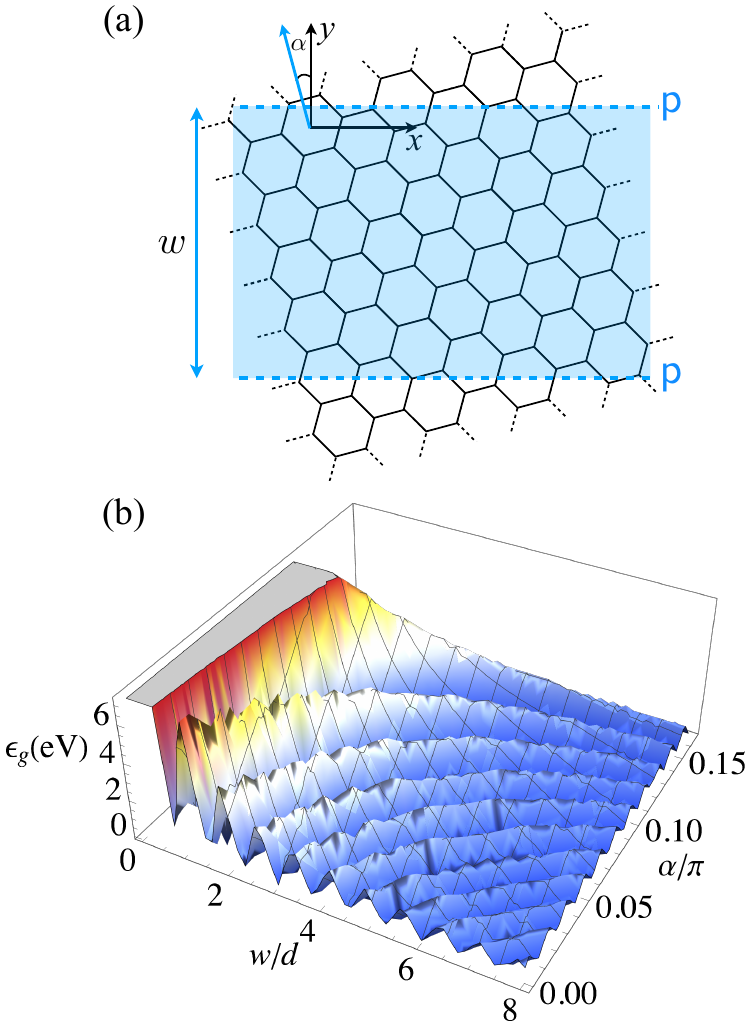}
\caption{(a) Schematic view for an intermediate realization of nanolane in which the honeycomb lattice is rotated by an angle of $\alpha$ (compare with Fig. \ref{GNLs} c).  (b) Energy bandgap, $\epsilon_\text{g}$, for an intermediate nanolane  is shown as a function of $w$ and $\alpha/\pi$ varying over the range $0\leq \alpha\leq \pi/6$.}
\label{im-e}
\end{figure} 

%%%%%%%%%%%%%%%%%%%%%%%%%%%%
\subsection{Intermediate nanolanes}
\label{subsec:int}

We now consider the intermediate case of nanolanes realized by rotating the honeycomb lattice with a tilt angle of $\alpha$; see Fig. \ref{im-e}a. In the example of a suspended graphene sheet over a rectangular trench, this realization can be viewed as the lattice being rotated relative to the edges of the trench. It is clear that the nanolane bandgap will periodically change with $\alpha$ in such a way that it reduces  to armchair and zigzag   states  for $\alpha=2j\pi/6$ and  $\alpha=(2j+1)\pi/6$, respectively, where  $j$ is a nonnegative integer.  Using skew coordinates, it is straightforward to obtain the generalized form of the wavevector $k_{2,n}$   as  
 \begin{equation} 
   k_{2,n}\!=\!\frac{\pi}{w}\left[n\!+\!\frac{8}{3}\!\left(\frac{w}{d}\right) \sin\left(\frac{\pi}6\!-\!\alpha\right) \right]. 
  \label{k-gen}
  \end{equation}
 where $0\leq \alpha\leq \pi/6$. As expected, for $\alpha=0$ and $\pi/6$, the armchair and zigzag   relations  \eqref{k-ac} and \eqref{k-zz} are recovered  (note that the zigzag nanolane in this case is aligned along the $x$-axis, compare  Figs. \ref{im-e}a and \ref{GNLs}c; hence, to recover Eq. \eqref{k-zz}, one further requires a relabeling $k_1\leftrightarrow k_2$).  The bandgap of intermediate nanolane is given by Eq. \eqref{ep-ac} provided one replaces $\Lambda$  with  
   \begin{equation}
 \Lambda \to \frac{8}{3}\!\left(\frac{w}{d}\right)\sin\left(\frac{\pi}6-\alpha\right)\,. 
\label{L1}
\end{equation}
 The result for the intermediate angle $\alpha=\pi/9$ is shown alongside the armchair and zigzag results in Fig. \ref{ac-e} (green curve). The energy bandgap of an intermediate nanolane as a function of  $w$ and $\alpha$ is shown in Fig. \ref{im-e}b.  As  seen, $\epsilon_\text{g}$  exhibits mainly armchair-like behaviors along both the $w$ and the $\alpha$ axes for $\alpha <\pi/6$. The bandgap here vanishes at widths  
 \begin{equation}
 w_\ast = \frac{3d}{8 \sin\left(\frac{\pi}{6}-\alpha\right)}\ell 
 \label{eq:w_int}
 \end{equation}
 for $\ell=1, 2, 3, \ldots$, which  identify the {\em metallic states} of an intermediate   nanolane. Equation \ref{eq:w_int} reduces to its previously mentioned  armchair  form when $\alpha=0$. 

Our foregoing results indicate that the predicted bandgaps of symmetric nanolanes   with arbitrary $\alpha$ and $w$ are {\em universal}; i.e., they remain independent of the boundary permeability $\textsf{p}$. 

%%%%%%%%%%%%%%%%%%%% 
\subsection{ Nanoribbons as impermeable nanolanes }
\label{subsec:GNR}

As noted before, our  model can be specialized to nanoribbons using the no-flux bag boundaries ($\mathsf{p}=0$) \cite{NR7,GQFT1,GQFT2,GQFT3,NR5,NR6} with displaced boundary lines by a total offset of $d$ to coincide with the respective armchair and zigzag lattice termination sites for nanoribbons \cite{NR4} (Figs. \ref{GNLs} b and d). Hence, for armchair nanoribbons, the permissible  wavevectors components $k_{2,n}$ are obtained from
    \begin{equation} 
   \sin [ (k_{2,n}-4\pi/3d) (w+d)] =0      \label{BB-AR}
  \end{equation}
where $w$ here is allowed to vary only over the discrete subset of values (subtypes) \cite{NR1,NR2,NR3,NR4} $w/d=3\ell\pm1$ and $3\ell$ ($\ell=1,2,3,\ldots$). For  $w/d=3\ell-1$, Eq. \eqref{BB-AR} yields  $k_{2,n}=n\pi/(w+d)$ reproducing the standard gapless states. For $w/d=3\ell$ and $3\ell+1$, Eq. \eqref{BB-AR} yields $k_{2,n}=n\pi/(w+d)+4\pi/(3d)$ and an energy band with no gapless states. In the implementation of bag boundaries for Dirac quasispinors in Refs. \cite{NR5,NR6}, the width of the confining area is taken as $w+d/2$ instead of the correct value of $w+d$, leading to  doubly degenerate states for $w/d=3\ell-1$ therein that differ from the tight-binding results \cite{NR1,NR2,NR3,NR4}. The  nanoribbon limit of our nanolane model through Eq. \eqref{BB-AR}, however, provides a significantly closer agreement with the tight-binding results. For instance, in comparison to the tight-binding results of Ref. \cite{NR4}, Ref.   \cite{NR5} produces relative deviations $|\Delta \epsilon_\textrm{g}|/\epsilon_\textrm{g}$ up to around $30\%$ (see Fig. 5 therein) when the width is reduced to $w=4d\simeq 1$~nm. The nanoribbon limit from our model gives relative deviations of $\epsilon_\textrm{g}\lesssim 4\%$ in comparison with Ref. \cite{NR4} for all $w$.  This can be seen by comparing the bandgap calculated from Eq. \eqref{BB-AR} (armchair nanoribbons), i.e.,  
 \begin{equation}
\frac{\epsilon_\textrm{g}}{\hbar \upsilon_\textrm{F}}=  \frac{2\pi}{1+w/d}\left|n+\frac43\left(1+\frac{w}{d}\right)\right|_\mathrm{min}
\label{e-D}
\end{equation}
for integer $n$, with the tight-binding expression (see appendix A of Ref. \cite{NR4}) 
\begin{equation}
\frac{\epsilon_\textrm{g}}{\hbar \upsilon_\textrm{F}}=\frac{4\sqrt{3}}{3}\left|1+2 \cos\left(\frac{r\pi}{N+1}\right)\right|_\mathrm{min}
\label{e-TB}
\end{equation}
where $r=1,2,\ldots,N$ and $N=2({w}/{d})+1$. For the narrowest gapped nanoribbon ($w=3d$),  Eqs. \eqref{e-TB} and \eqref{e-D} give $\epsilon_\textrm{g}\simeq 0.54\,\hbar \upsilon_\textrm{F}$ and $0.52\,\hbar \upsilon_\textrm{F}$, respectively, at the relative 4\% difference noted above. Our results thus indicate that the continuum modeling can remain relatively  accurate down to just a couple of lattice spacings. This contrasts the conclusion in Ref. \cite{NR5} that attributes the reported sizable deviations at small widths to the shortcomings of the continuum model. Our results and those of Refs. \cite{NR5,NR6} and the tight-binding models asymptotically converge at large widths. 

For the limiting case of zigzag nanoribbons, our model gives a simpler form of Eq. \eqref{BB-AR} as $\sin [k_{1,n}(w+d)] =0$ which produces a gapless energy band regardless of the choice of $w$.   This agrees with the tight-binding results \cite{NR1,NR2,NR3,NR4}. 
% (the solutions for $k_{\myperp,n}$ reproduce the analytical limit of large $w$ from the tight-binding calculations \cite{NR4}; i.e., $k_{\myperp,n}\simeq n\pi/w$). 
The limiting zigzag nanoribbon obtained from our nanolane model can however be distinguished from  zigzag nanoribbons in the following two aspects. First, unlike the tight-binding results, the model here  gives a  closed-form expression for the wavevector as in Eq. \eqref{k-zz}. Second, it does not possess any edge-states. The latter is expected, as the nanolanes in our model produce no physical lattice termination edges. The edge-states also appear to signify the  Dirichlet boundary conditions  that are imposed asymmetrically relative to the two different honeycomb sublattices of graphene sheet  in both the tight-binding \cite{NR4,NR7} and previous massless Dirac formulations \cite{NR5} but not in our model.  

%%%%%%%%%%%%%%%%%%%%%%%%%%%%%%%%%%%%%%%%%%%%%%
\section{Concluding remarks}
\label{sec:conc}

We have proposed  a generalized form of bag boundary conditions to address confinement of massless Dirac quasispinors within permeable quasi-1D areas, introduced as nanolanes,  across an unbounded  monolayer graphene sheet. The problem is formulated within a continuum field-theoretic approach that captures the essential physics of low-energy electronic excitations over the honeycomb lattice  \cite{GQFTrev1,GQFTrev2,GQFT3,GQFT4}. The model nanolanes are obtained by (1) introducing any pair of parallel straight lines at arbitrary (hence, continuously  varying) distance from one another over an unbounded lattice, and (2) assigning finite permeabilities to quasispinor currents at these boundaries. These constructions not only preserve formal consistency of the model with the underlying Dirac equation (as opposed to, e.g., Dirichlet boundary conditions) but they also allow  for the nanolane properties (here, specifically, the bandgap) to be {\em tuned} via continuous variables. The latter include the nanolane width $w$ and the tilt angle $\alpha$ that  can itself be varied continuously to cover all intermediate nanolane configurations between armchair ($\alpha=0$) and  zigzag ($\alpha=\pi/6$) states. 

Armchair nanolanes are found to exhibit  both metalized and insulating states depending on $w$. By contrast, zigzag nanolanes are found to be always metalized. The quantitative predictions of our model remain accessible to experimental verification. Specifically, our results  predict sizable bandgaps,  e.g., up to $\simeq 2$~eV for a nanometer-wide nanolane, larger than those obtained from tight-binding calculations in the case of nanoribbons ($\lesssim 1$~eV)  \cite{NR1,NR2,NR3,NR4}. 

In the special limit that corresponds to graphene nanoribbons (i.e., impermeable nanolanes with widths varying only over a discrete set of admissible values), our approach recovers  the tight-binding bandgaps   \cite{NR1,NR2,NR3,NR4}  to within relative errors of only a few percents even at small nanoribbon widths of a couple lattice spacings. In fact, the bandgap in our approach is derived more straightforwardly and in only a few steps of calculation relative to the tight-binding approach \cite{NR4}. In the limit of nanoribbons, we also clarify the source of discrepancies between previous implementation of no-flux bag boundaries   \cite{NR5} and the tight-binding results (Section \ref{subsec:GNR}). 

 Our analysis assumes the validity of continuum-limit quasispinor solutions \eqref{psi} (and evidently also the Dirac dispersion relation, etc.) in the nanolane geometry at hand. These kinds of solutions have previously been used in the study of nanoribbons (see, e.g.,  Refs. \cite{GR1,GQFT2,GQFT3,NR5,NR6}). The close agreements that these works and (to an improved extent) our current work establish between the continuum and tight-binding solutions give direct support to the applicability of continuum approach in  analyzing quasispinor properties of nanoribbons. A similar comparison between continuum results and the tight-binding ones still remains to be conducted  in the case of nanolanes. Nevertheless, we expect deviations from the continuum limit to be weaker in nanolanes as compared with nanoribbons. This is because nanolanes involve a weaker confinement of quasispinors that are  allowed to partially permeate to the unbounded graphene regions outside the nanolane. 

 Our study can also facilitate theoretical modeling of electronic confinement in graphene in a wider range of  realistic situations where the confinement is expected to be nonideal. Possible realizations of boundary lines in our model nanolanes may include arrays of adatoms deposited over a graphene sheet or, for a graphene sheet suspended over a rectangular trench, the parallel contact lines created between  the sheet and the edges of the trench. In the latter example, the contact  lines across the edges of the trench can generally be incommensurate  with armchair/zigzag edges of a nanoribbon and/or may not necessarily be  impermeable to quasispinor currents, making the nanolane construction a more viable model for the purpose (although further generalizations of the model, e.g., by incorporation of substrate effects,  may need to be considered as well).
%In both these examples the boundary lines may not necessarily be taken as lattice termination points as in nanoribbons. 
%Such a setup can be considered as a prototypical model for nanoelectromechanical  systems and nano-scale resonators that can be actuated by  some external forces to vibrate in  particular resonance modes  \cite{NEMS1,NEMS2,NEMS3}. 
While the proposed nanolanes can be viewed  as  a promising toy-model to study electronic control in  graphene-based nanosystems \cite{Gtec1,Gtec2,Gtec3,Gtec4,Gtec5}, it is also important to analyze its predictions using other more comprehensive techniques such as first-principle calculations \cite{FP1,FP2,FP3} which, for nanoribbons, predict sizable corrections to the tight-binding results in closer agreement with experiments \cite{exp1,exp2,exp3,exp4,exp5}. 

 As possible extensions of our analysis, one may consider model nanolanes with non-identical boundary permeabilities and/or boundary lines of finite width. In these or other possible generalizations of the problem, the explicit form (e.g., wavevector-dependence) of the boundary permeabilities might become relevant and require further exploration. Such effects are manifestly absent in the present context with symmetric nanolanes that exhibit universal (permeability-independent) bandgaps.
 
Finally, we note that the special case of no-flux bag boundaries can also be implemented in its indirect form \cite{GQFT1,GQFT2} as $\icomplex \gamma^a n_a \psi=\psi$ while we have used its direct form (i.e., Eq. \eqref{BB} with $\mathsf{p}=0$). Our inspections  indicate that the two formulations may not be necessarily equivalent and it is the latter that accurately reproduces the tight-binding results (in fact, the relevant nanoribbon subtypes $w/d=3\ell\pm1, 3\ell$ do not emerge in the former case \cite{GQFT1,GQFT2}). In the nanolane context,  the distinction between the two formulations of bag boundaries  is even more apparent, as  the boundaries exhibit finite permeabilities to quasispinor currents. 
   
\section*{Conflict of Interest}
The authors have no conflicts to disclose.

\section*{Data Availability}
The data that support the findings of this study are available within the article.

%%%%%%%%%%%%%%%%%%%%%%%%%%%%%%%%%%%%%%%%%%%%%%

%%%%%%%%%%%%%%%%%%%%
\end{document}